\documentclass[11pt,twoside]{article}
\usepackage{./asp2014}
\usepackage{xcolor}
\usepackage{hyperref}
\newcommand{\CM}[1]{\textcolor{black}{#1}}
\aspSuppressVolSlug
\resetcounters
\usepackage{titlesec}
\titlespacing*{\section}{0pt}{1.0\baselineskip}{0.7\baselineskip}

\begin{document}

\title{The ARCHES project}
\author{C.Motch,$^1$ F. Carrera,$^2$, F. Genova,$^1$ F. Jim\'enez-Esteban,$^3$ M. L\'opez,$^3$ L. Michel,$^1$ B. Mingo,$^4$ A. Mints,$^5$ A. Nebot,$^1$ F.-X. Pineau,$^1$ S. Rosen,$^4$ E. Sanchez,$^1$ A. Schwope,$^5$ E.Solano,$^3$ and M.Watson $^4$
\affil{$^1$Observatoire Astronomique, Strasbourg, France \email{}}
\affil{$^2$Instituto de F\'{i}sica de Cantabria, (CSIC-UC), Santander, Spain \email{}}
\affil{$^3$Centro de Astrobiolog\'{i}a, Villanueva de la Ca\~{n}ada, Spain \email{}}
\affil{$^4$University of Leicester, Leicester, UK \email{}}
\affil{$^5$Leibniz-Institut f\"{u}r Astrophysik, Potsdam, Germany \email{}}
}

% This section is for ADS Processing.  There must be one line per author.

\paperauthor{Christian Motch}{christian.motch@astro.unistra.fr}{0000-0002-4298-5624}{Observatoire
astronomique de Strasbourg, Université de Strasbourg, CNRS, UMR
7550}{Observatoire Astronomique de Strasbourg}{Strasbourg}{N/A}{F-67000}{France}

\paperauthor{Francisco J. Carrera}{carreraf@ifca.unican.es}{0000-0003-2135-9023}{Instituto de Fisica de Cantabria}{(CSIC-Universidad de Cantabria)}{Santander}{}{E-39005}{Spain}

\paperauthor{Francoise Genova}{francoise.genova@astro.unistra.fr}{0000-0002-6318-5028}{Observatoire
astronomique de Strasbourg, Université de Strasbourg, CNRS, UMR
7550}{Observatoire Astronomique de Strasbourg}{Strasbourg}{N/A}{F-67000}{France}

\paperauthor{Francisco Jimenez-Esteban}{fran.jimenez-esteban@cab.inta-csic.es}{}
{Centro de Astrobiolog\'{\i}a (INTA-CSIC)}{Departamento de Astrof\'{\i}sica}{Villanueva de la Ca\~nada}{}{E-28691}{Spain}

\paperauthor{Mauro L\'opez}{mauro@cab.inta-csic.es}{}
{Centro de Astrobiolog\'{\i}a (INTA-CSIC)}{Departamento de Astrof\'{\i}sica}{Villanueva de la Ca\~nada}{}{E-28691}{Spain}

\paperauthor{Laurent Michel}{laurent.michel@astro.unistra.fr}{}{Observatoire
astronomique de Strasbourg, Université de Strasbourg, CNRS, UMR
7550}{Observatoire Astronomique de Strasbourg}{Strasbourg}{N/A}{F-67000}{France}

\paperauthor{Beatriz Mingo}{bm188@leicester.ac.uk}{}{X-ray And Observational Astronomy Group
Department of Physics and Astronomy}{University of Leicester}{Leicester}{N/A}{LE1 7RH}{UK}

\paperauthor{Alexey Mints}{amints@aip.de}{0000-0002-8440-1455}{Leibniz-Institut für Astrophysik Potsdam}{AIP}{Potsdam}{N/A}{D-14482}{Germany}

\paperauthor{Ada Nebot}{ada.nebot@astro.unistra.fr}{}{Observatoire
astronomique de Strasbourg, Université de Strasbourg, CNRS, UMR
7550}{Observatoire Astronomique de Strasbourg}{Strasbourg}{N/A}{F-67000}{France}

\paperauthor{Fran\c{c}ois-Xavier Pineau}{francois-xavier.pineau@astro.unistra.fr}{}{Observatoire
astronomique de Strasbourg, Université de Strasbourg, CNRS, UMR
7550}{Observatoire Astronomique de Strasbourg}{Strasbourg}{N/A}{F-67000}{France}

\paperauthor{Simon Rosen}{srr11@leicester.ac.uk}{}{X-ray And Observational Astronomy Group
Department of Physics and Astronomy}{University of Leicester}{Leicester}{N/A}{LE1 7RH}{UK}

\paperauthor{Estrella Sanchez}{estrella.sanchez-ayaso@astro.unistra.fr}{}{Observatoire
astronomique de Strasbourg, Université de Strasbourg, CNRS, UMR
7550}{Observatoire Astronomique de Strasbourg}{Strasbourg}{N/A}{F-67000}{France}

\paperauthor{Axel Schwope}{ASchwope@aip.de}{0000-0003-3441-9355}{Leibniz-Institut für Astrophysik Potsdam}{AIP}{Potsdam}{N/A}{D-14482}{Germany}

\paperauthor{Enrique Solano}{esm@cab.inta-csic.es}{0000-0003-1885-5130}
{Centro de Astrobiolog\'{\i}a (INTA-CSIC)}{Departamento de Astrof\'{\i}sica}{Villanueva de la Ca\~nada}{}{E-28691}{Spain}

\paperauthor{Mike Watson}{mgw@leicester.ac.uk}{}{X-ray And Observational Astronomy Group
Department of Physics and Astronomy}{University of Leicester}{Leicester}{N/A}{LE1 7RH}{UK}

\begin{abstract}
ARCHES (Astronomical Resource Cross-matching for High Energy Studies) is a FP7-Space funded project whose aim is to provide the international astronomical community with well-characterised multi-wavelength data in the form of spectral energy distributions (SEDs) for large samples of objects extracted from the 3XMM DR5 X-ray catalogue of serendipitous sources. The project has developed new tools implementing fully probabilistic simultaneous cross-correlation of several catalogues for unresolved sources and a multi-wavelength finder for clusters of galaxies for extended sources. These enhanced resources have been tested in the framework of several science cases.
\end{abstract}

\section{Introduction}

The opening up of new observing windows in the electromagnetic spectrum and the development of large area detectors has revolutionised observational astronomy during the last decades. In particular, the high-energy window to the Universe has strongly benefited from the availability of major new facilities such as the XMM-Newton and Chandra space observatories which routinely observe the X-ray sky and generate unprecedentedly large catalogues of X-ray sources. 
Most recent observational studies acknowledge the scientific importance of the exponentially increasing collection of archival data made available and include a multi-wavelength approach from the very beginning.  

However, the availability and the quality assessment of multi-wavelength data can be an obstacle. Another difficulty lies in establishing the correct identification of a given astrophysical object over a large range of wavelengths and with significantly varying astrometric quality among different catalogues. This last problem is particularly crucial for the identification of XMM-Newton sources which due to the relatively large point spread function of the X-ray telescope  suffer from source blending and lower astrometric accuracy \citep[see e.g.][]{2009A&A...493..339W}.

Taking these considerations into account, the FP7-Space ARCHES project was started with the aim to create qualified spectral energy distributions (SEDs) over a large wavelength range from X-rays to radio for all good quality unresolved X-ray sources in the 3XMM DR5. For that purpose, ARCHES has selected the most relevant archival catalogues with the best sky coverage, homogeneity, and astrometric and photometric quality and developed an original multi-catalogue statistical cross-correlation tool. For extended sources, ARCHES has designed an integrated cluster finder that is able to automatically search for groups of galaxies at consistent redshifts in optical and near-infrared catalogues while providing the statistical significance of the candidate cluster.

ARCHES products and tools are made available to the general community via several channels\footnote{\CM{see the ARCHES web site at \href{www.arches-fp7.eu}{www.arches-fp7.eu}}}. The results of the multi-catalogue cross-correlations, the SEDs and the list of candidate clusters of galaxies are distributed through specific interfaces\footnote{\CM{SSC catalogue server at Strasbourg  \href{http://xcatdb.unistra.fr/}{http://xcatdb.unistra.fr/}}} and as flat files by the CDS Vizier server\footnote{\CM{\href{http://vizier.u-strasbg.fr/viz-bin/VizieR?-source=IX/48}{http://vizier.u-strasbg.fr/viz-bin/VizieR?-source=IX/48}}}. The cross-correlation tool is available on a dedicated machine\footnote{\CM{\href{http://serendib.unistra.fr/ARCHESWebService/index.html}{http://serendib.unistra.fr/ARCHESWebService/index.html}}} and will eventually be merged in the CDS services. ARCHES has also created outreach facilities, in particular the Arches Walker (see L. Michel et al. 2015, this volume). 

\section{ARCHES tools and products}

\subsection{An enhanced 3XMM DR5 catalogue}

The first step of the project has been to create an enhanced version of the 3XMM DR5 source catalogue (3XMMe) by gathering best quality 3XMM detections \citep{Rosen2016short}. For instance, sources located in a subset of 19 mosaic-mode observations where issues in original data files led to uncertain detections were discarded. We also rejected 568 fields with high background levels assessed through visual screening. Sources acquired at off-axis angles $>$ 12\arcmin, with exposure times $<$ 5 ks, with only windowed MOS modes or for which the {\it catcorr} boresight correction failed were also rejected as well as a few sources that were not recognised as due to bad pixels by the production pipeline. Improved parameters were derived from the final set of cleaned detections for each unique source. In addition to cleaning, unique sources were identified as being suitable candidates for use in the specific science themes of the ARCHES project (e.g. criteria on galactic latitude, field content, etc.). The currently released version 2.0 of the 3XMMe contains 285205 detections of 219788 unique sources from 4802 observations and can be downloaded from the ARCHES web site and from VizieR\footnote{\CM{\href{http://vizier.u-strasbg.fr/viz-bin/VizieR?-source=IX/47}{http://vizier.u-strasbg.fr/viz-bin/VizieR?-source=IX/47}}}.

\subsection{The ARCHES advanced cross-correlation tool}

We established the foundations for probabilistic multi-catalogue cross-matches of unresolved sources. The selection of candidates is based on chi-square hypothesis tests using elliptical error definitions. Bayesian probabilities of association are computed taking into account all possible combinations of association or non associations in the various catalogues in such a way that the sum of all probabilities is equal to one. Priors are derived from local source densities and include the effects of the chi-square selection criterion. Drawing on these theoretical advances, ARCHES has developed a generic and flexible tool able to cross-correlate, in a single pass, an arbitrary large number of catalogues extracted from the Vizier CDS service while providing probabilities for each combination of association/non association \citep{Pineau2016}. Importantly, the algorithm is flexible enough to handle a variety of cross-correlation scenarios. For instance, one can search all sources in a master catalogue for all possible counterparts in at least one or several other catalogues. In the framework of ARCHES, the cross-correlation tool was used with the 3XMMe catalogue as master catalogue. Alternatively, the tool can also be used to extract all sources common to say, three catalogues, or all sources common to two catalogues and necessarily absent in a third one, etc.. Although there is no theoretical limitation to the number of different catalogues that can be processed in a single pass, due to the geometrically increasing number of possible combinations, the number of catalogues to cross-match can hardly exceed eight. The ARCHES cross-match tool contains about 34,000 lines of Java code and can be accessed using an http API that submit scripts on a dedicated machine. In the relatively near future a Web interface will provide a more simple use of the tool based on a set of pre-defined cross-correlation scenarios. 

\subsection{The ARCHES cross-matched catalogues and the SEDs}

A large effort has been invested into the selection of the archival catalogues based on which the SEDs of all 3XMMe sources would be assembled. Emphasis was primarily put on the quality (e.g. uniform photometric and astrometric  properties and well controlled source samples) and on the source classification value (sky coverage, overlap with 3XMMe and uniqueness). Starting from a list of over 200 candidate catalogues, a total of 12 catalogues were finally selected (GALEX, UCAC, SDSS, V/IPHAS, 2MASS, UKIDSS, WISE, GLIMPSE, AKARI/FIS and a merge of FIRST, NVSS and SUMS for the radio band). Systematic astrometric uncertainties were carefully assessed, as these quantities may in fact dominate the error budget in many cases.  Owing to their lower astrometric quality, cross-correlation with AKARI and radio catalogues was made in a second step. ARCHES produced three main cross-correlated catalogues, one devoted to the Galactic plane (including GLIMPSE) and two for general purpose use (one with 2MASS and one with UKIDSS). These tables contain a summary of the most important physical quantities extracted from the archival catalogues as well as a description of the five most probable combinations of candidates. We use the SVO Filter Profile Service \citep{2013hsa7.conf..953R} to convert magnitudes into fluxes. These catalogues are downloadable from the ARCHES web site. They can also be browsed and data mined using the ArchesDb\footnote{\href{http://xcatdb.unistra.fr/3xmmdr5/archesindex.html?mode=catalog}{http://xcatdb.unistra.fr/3xmmdr5/archesindex.html?mode=catalog}}, an interface inherited from the 3XMM XcatDb \citep{2015ASPC..495..173M} and offering the same query and data display capabilities. SED plots and multi-catalogue finding charts are also available in the ArchesDb. The SED VO-compliant archive\footnote{\href{http://sdc.cab.inta-csic.es/arches/}{http://sdc.cab.inta-csic.es/arches/}} provides access to spectral energy distributions generated from the cross-correlated catalogues. The system allows one to easily query, visualise, download and send SEDs to VO-tools via SAMP protocol.

\subsection{The ARCHES integrated cluster finder}

Most high Galactic latitude extended X-ray sources are known to be associated with clusters of galaxies. However, confirming their identification requires finding, on the line of sight, an over-density of galaxies at a common redshift. ARCHES has thus designed a tool searching for clusters of galaxies and estimating their parameters (redshift, size and X-ray parameters) by combining XMM-Newton and ground- and space-based multi-wavelength catalogues. The Integrated Cluster Finder tool (ICF) \citep{mints2015} actually works at any place in the sky covered by SDSS, UKIDSS and WISE. Cosmological redshift is measured using colour-redshift relations. The spectral energy distribution of member galaxies and of the central dominant galaxy, supplemented by optical spectra, if available, provide an estimate of the redshift. At redshift higher than $z\,\geq\,0.55$ redshift determination is improved by including an infrared band. The ARCHES cross-correlation tool is used to cross-match optical and infrared sources. The ICF uses a redMaPPer-like algorithm \citep{2014ApJ...785..104} in which the multiplicity function $\lambda(z)$ measures the probability of finding a genuine cluster at a given redshift. Mainly written in Python, most computing intensive parts are coded in Fortan. The ICF has been successfully tested against the samples of clusters in the literature. Applied to the DR5 version of the 3XMMe catalogue, the ICF identifies about 500 candidate clusters. Plots of $\lambda(z)$ and finding charts showing SDSS images with candidate cluster members overlaid with XMM X-ray contours are generated for each extended 3XMMe source. The ICF tool is accessible from a dedicated server located in Strasbourg while the catalogue and its associated graphical material can be browsed in the ArchesDb.

\section{ARCHES science cases}

ARCHES products are at the basis of five test science cases, the results of which will be reported elsewhere; i) a study of obscured accretion at optical, X-ray and mid-infrared wavelengths and of the effects of different spectral bands for selecting AGN, ii) a study of jet mechanisms and accretion in AGN based on a large sample ($\sim$1000\,sources) of radio-loud AGN and star forming galaxies selected and characterised using 3XMMe, radio (FIRST+NVSS) and mid-IR (WISE), iii) a follow-up study of XMM-Newton discovered clusters and a comparison with SZ-based catalogues, iv) the identification, characterisation and modelling of the X-ray emitting stellar population, and v) a multi-wavelength study of debris discs around late-type stars. 

\bibliography{astrophP077}  % For BibTex

\end{document}